\pdfoutput=1 
\documentclass{JINST}

\def \km3 { KM3NeT }

\title{The Central Logic Board and its auxiliary boards
for the optical module of the \km3 detector}

\author{S. Biagi$^a$ and
A. Orzelli$^b$\thanks{Corresponding author.}\, on 
behalf of the \km3 Collaboration\\
\llap{$^a$}University of  Bologna and INFN,\\
  Viale Berti Pichat 6/2 - 40127 - Bologna, Italy\\
E-mail: \email{simone.biagi@bo.infn.it}\\
\llap{$^b$}INFN Genova,\\
  Via Dodecaneso 33 - 16146 - Genova, Italy\\
E-mail: \email{aorzelli@ge.infn.it}}

\abstract{
The \km3 neutrino telescope will be composed of many optical modules, 
each of them containing 31~(3") photomultipliers, connected to a Central 
Logic Board. The Central Logic Board integrates Time to Digital Converters 
that measure Time over Threshold of the photomultipliers signals while 
White Rabbit is used for the optical modules time synchronization. 
Auxiliary boards have also been designed and built in order to test and 
extend the performance of the Central Logic Board. The Central Logic Board, 
as well as the auxiliary boards, will be presented by focusing on the 
design consideration, prototyping issues and tests.
}


\begin{document}

\section{Introduction}

\km3 is a deep-sea research infrastructure, which will host a 
neutrino telescope with a volume of several cubic kilometres at 
the bottom of the Mediterranean Sea \cite{km3net}. It will be composed by 
thousands of Digital Optical Modules (DOMs), consisting of a glass 
sphere containing 31~(3") PhotoMultiplier Tubes (PMTs) for the detection 
of the Cherenkov light induced by charged particles produced by 
the interaction of   neutrinos with matter inside or in the vicinity 
of the \km3 detector \cite{dom,ppmdom}. A group of 18~DOMs distributed over a 700~m 
mooring line constitutes a Detection Unit  of the telescope.

The signal acquired by each photomultiplier is sent to a Time over Threshold (ToT) 
discriminator to fed the Time to Digital Converter (TDC) \cite{tdc} which is 
part of a Central Logic Board (CLB) based on the Kintex 7 FPGA.
The TDC resolution is 1 ns and the White Rabbit technology \cite{wr} is used to guarantee 
time synchronization at the level of 1 ns between each DOM.

Additional peripheral devices are connected to the CLB, in order to keep 
track of both the environmental conditions (temperature, humidity), the DOM 
orientation (yaw, pitch, roll) and its position; some of them are embedded 
on the board, such as the temperature~\&~humidity sensor and the tiltmeter~\&~compass, 
while others are plugged to the CLB by using connectors, such as the acoustic 
devices (Piezo or Hydrophone) and the Nanobeacon (a LED device). 
Custom boards have also been designed and produced to test the performances 
of the CLB or to extend its functionalities. 

All the incoming data are collected by the CLB and sent to shore using a dedicated 
optical network. Two Lattice Microcontrollers 32-bit (LM32)  \cite{lm32} are implemented in the CLB FPGA, 
one for the White Rabbit and the other for the instrumentation management.

Two batches of prototypes have been produced and are now operational in the 
laboratories of the \km3 Collaboration; 60 pieces were ordered to equip the first 
two \km3 Detection Units which will be deployed during 2015.
A prototype Detection Unit \cite{ppmdu} has been deployed in May 2014 at the KM3NeT-It 
installation site 100~km~SE off shore of Capo Passero, Sicily, and it is continuously taking 
data since its connection to the on-shore station.

\section{The Central Logic Board}

\begin{figure}[tbp] 
\centering
\includegraphics[width=.7\textwidth]{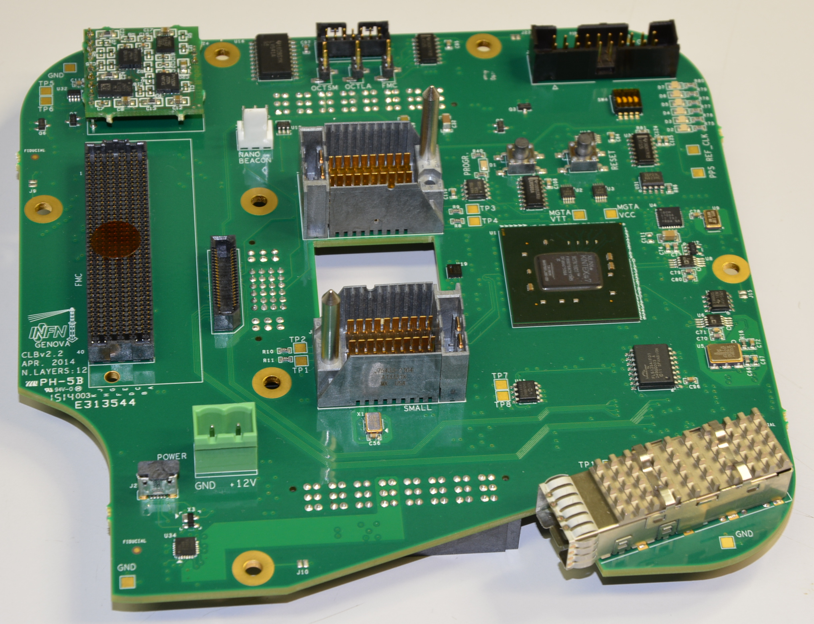}
\caption{The Central Logic Board.}
\label{fig:clb}
\end{figure}

\begin{figure}[tbp] 
\centering
\includegraphics[width=.45\textwidth]{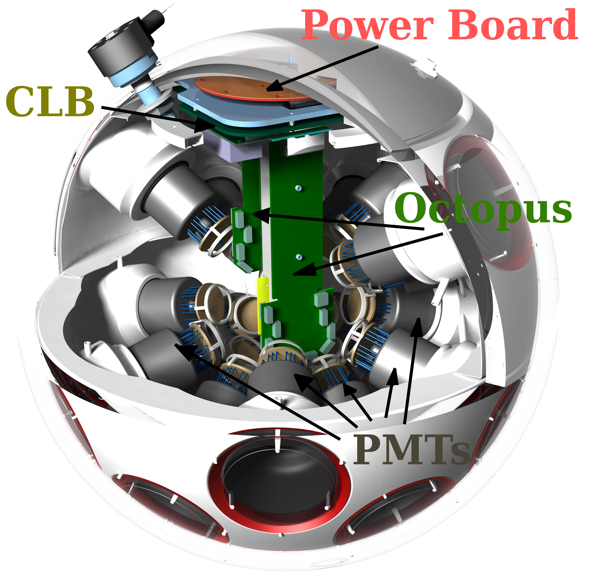}
\caption{Schematic view of a \km3 DOM.}
\label{fig:dom}
\end{figure}

The CLB, showed in figure~\ref{fig:clb}, is the main logic board of 
the \km3 DOM (see figure~\ref{fig:dom}). 
It is directly connected with
a Power Conversion  Board which provides all the needed rails for the board (1.0 V,
1.8 V, 2.5 V, 3.3 V, 5.0 V).
The purpose of the CLB is to manage all the instrumentation present inside the
DOM, acquire incoming data  from the 31 PMTs and from the  acoustic device, pack the data
into User Datagram Protocol (UDP) packets and send them through the optical line. A time synchronization
of 1~ns level between all the DOMs is achieved by the usage of White Rabbit.

\begin{figure}[tbp] 
\centering
\includegraphics[width=\textwidth]{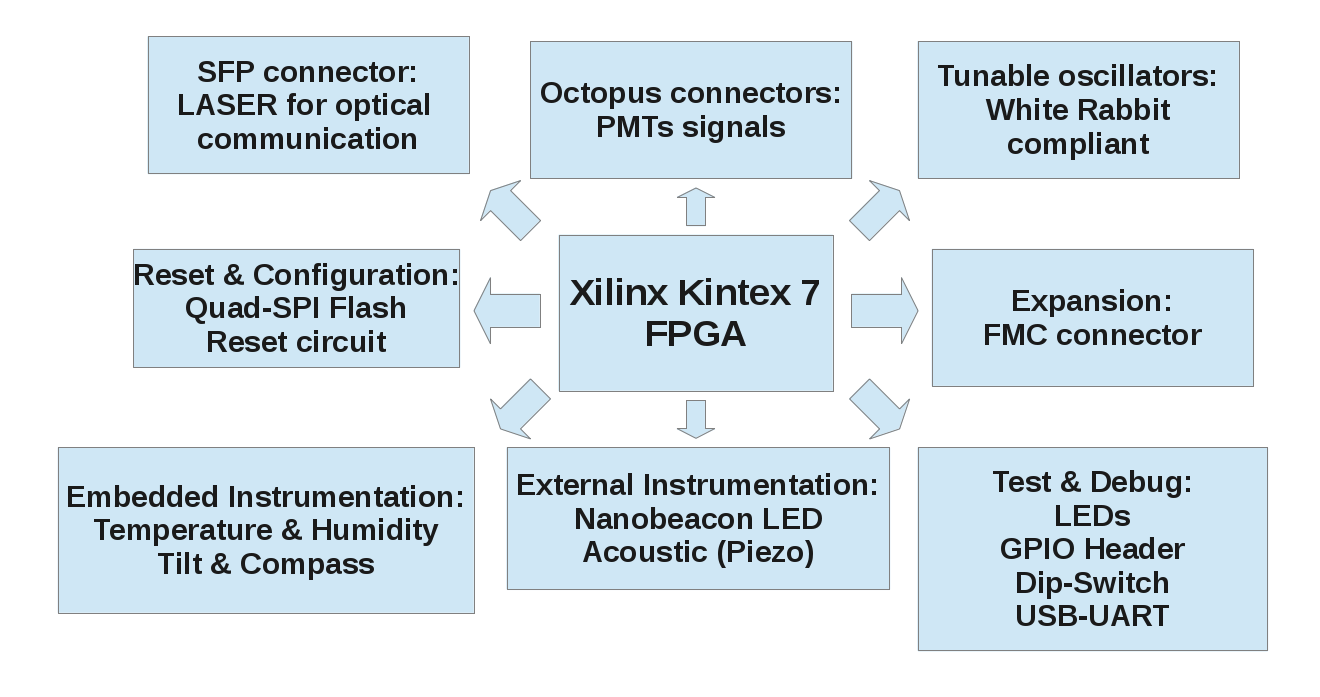}
\caption{Schematic block of the Central Logic Board.}
\label{fig:block}
\end{figure}

A schematic block of the CLB is shown in figure~\ref{fig:block}. It is
based on a Xilinx Kintex 7 FPGA, which is connected to all the rest of the
board, including in particular:
\begin{itemize}
	\item 1 Small Form-factor Pluggable (SFP) connector for the optical communication
	\item 2 custom Octopus connectors for the PMTs signals
	\item Tunable oscillators (White Rabbit compliant)
	\item Embedded instrumentations (temperature and humidity sensor, tilt~\&~compass)
	\item External connectors (for Nanobeacon LED, acoustic device, expansion boards)
\end{itemize}

\section{The firmware}

\begin{figure}[tbp] 
\centering
\includegraphics[width=\textwidth]{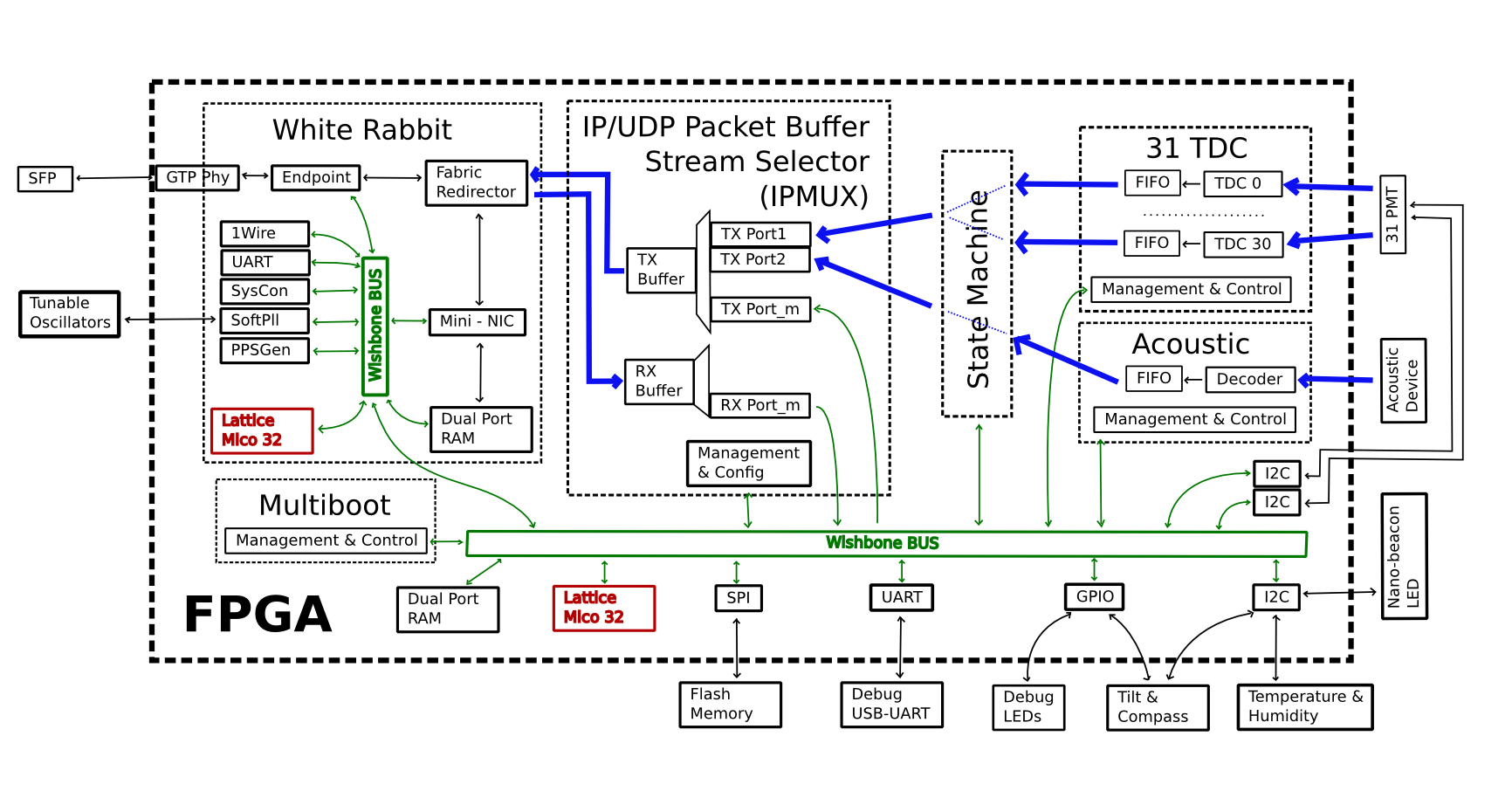}
\caption{Schematic block of the Central Logic Board firmware.}
\label{fig:fw}
\end{figure}

The firmware of the CLB (see figure~\ref{fig:fw}) is based on two 
LM32; Wishbone buses  are used to interconnect the microcontrollers with all the other
Intellectual Property cores.
Part of the firmware is dedicated to the White Rabbit (left in the figure), which
directly manages the tunable oscillators and the optical link, in order to achieve a
time synchronization of sub-nanosecond level with the Grand Master clock of the 
on-shore station. All the rest of the board is managed by the second microcontroller, which 
has access to all the drivers (SPI, UART, GPIO and I$^2$C) needed for the instrumentation,
the configuration and the test of the board.

A multiboot module allows the selection of a different image from where to boot the
FPGA, and the presence of a Golden Image as a fallback solution, will make possible to
remotely update the firmware after the deployment of the DOM.

An acoustic module is dedicated to the decoding of the incoming AES3 formatted stream, 
while 31~TDCs are responsible to record the timestamp and the width (with 1~ns of 
resolution) of hits incoming
from the ToT  signals of the PMTs. All the data, together with some
other slow control monitoring information (as temperature, humidity, tilt~\&~compass, 
currents, \ldots) are put in UDP packets by a state machine, and sent to the IPMUX module, 
the responsible for the management of the transmission and reception of packets to/from the
on-shore station.

\section{The layout}

The CLB layout is composed of 12 layer, with:
\begin{itemize}
	\item Six Signal layers including top and bottom
	\item Two Power planes
	\item Four ground planes
\end{itemize}

The layers are symmetrically disposed around the 2 Alimentation layers; ground planes 
are positioned at the sides of the signal layers to have better signal integrity, as well
as a limited number of vias was used. Particular attention was put on the differential
pairs routing keeping the time difference of less than 100~ps between different
PMT signals, and less than 20~ps between clock signals.

A reliability analysis using the FIDES method \cite{fides} was performed, showing an 
estimated risk for failure 
of less than 10~\% after 15 years.

Several signal integrity simulation were performed on different signals on the board, 
always showing a good level discrimination.

\section{Tests}

Several tests have been done on the CLB by the electronic group of the \km3 collaboration.
The tests covered all the functionality of the board, and where performed separately,
by focusing on the dedicated hardware and firmware section. Currently 
integration tests with a full assembled DOM are undergoing,
and all the  functionalities of the DOM are enabled.

The first and most important tests are about the time synchronization
and the optical communication link; the White Rabbit section of the firmware and  
hardware were tested by connecting a CLB with a Simple PCIe FMC carrier  card. 
The two boards used the 
White Rabbit Precision Time Protocol to achieve a sub-nanosecond level time synchronization, showed on 
the oscilloscope by using a pulse per second output signal from both the two boards. Other 
tests, with more boards involved and White Rabbit switches used to simulate the final
topology of the \km3 network followed.

The second  test regarded the PMT signals acquisition. 
Using a pattern generator plugged on the Octopus connector of the CLB, a regular pattern
was sent to simulate  all the 31 signals and the TDCs in the FPGA firmware were tested, together with
the UDP packet building and sending systems. Results were analyzed by software applications
on PC directly connected to the optical line of the CLB. Then more tests followed, 
by using real PMTs, connected through the Octopus boards and putted in a dark box. Hit rates
were analyzed in different conditions of high voltage and threshold settings for the ToT  measurement.

Then the acoustic module was tested by plugging a prototype of the final acoustic device
to the CLB and sending the decoded data to a PC using the UDP packet transmission; the multiboot
module was tested as well, and all the instrumentation on the board were tested by using
a shell command interface with the USB-UART connection implemented in the software embedded, 
before redirecting all the data output to the optical line.

A thermal analysis was also performed on the board, showing two hot regions near the FPGA and 
near the SFP connector, having in both case a maximum temperature of about 40~$^\circ$C. A
second measurement has been performed by inserting the CLB and the Power Board in the
final mechanical structure, which act as a dissipator, and the temperature on 
the CLB was found decreased by about 10~$~^\circ$C. Only the core of the FPGA reached 
50~$^\circ$C.

Finally an Electromagnetic Interference analysis was performed on the CLB with the Power Board 
plugged below it. The results showed the presence of two hot spots in the center of the boards, 
due to DC-DC converters on the Power Board; a second measurement was done by inserting a metal 
plane between the two boards, as it will be in the final design, and the two spots were not 
present anymore.
A third hot spot was found on the top of the CLB, due to a test signal that will not be
active during normal operation.

\section{Auxiliary boards}

The \km3 detector foresees to use the CLB not only inside the DOM, but
also in the Detection Unit Base and in the Calibration Unit. Here the CLB functionalities 
will be slightly different from the standard one, for example  no PMT measurements will be
done, but the control of other instrumentation will be required.

\begin{figure}[tbp] 
\centering
\includegraphics[width=.4\textwidth]{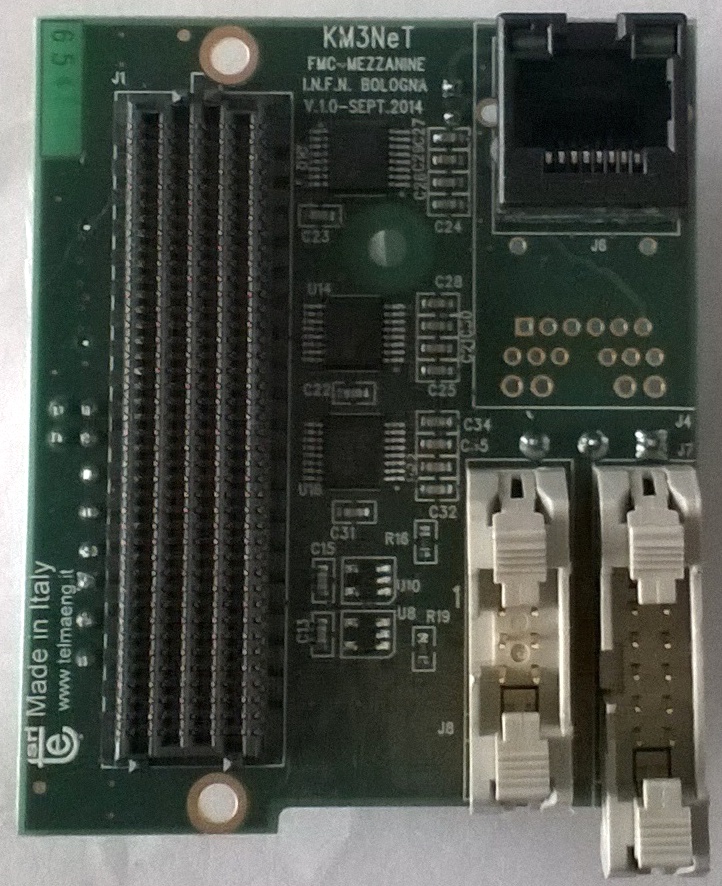}
\caption{ FMC Expansion Board for Detection Unit Base and Calibration Unit.}
\label{fig:fmc}
\end{figure}

In order to make all the needed connections available to the CLB, 
a  FPGA Mezzanine Card (FMC)  Expansion Board (see figure~\ref{fig:fmc}) has been designed and 
produced; in particular, small dimensions are required to fit in the mechanical constraints of the base modules.
The Expansion Board provides the following additional ports:
\begin{itemize}
	\item 3 additional RS232 ports
	\item 2  trigger lines with the RS485 standard
	\item 1 additional acoustic interface, using an RJ45 connector
\end{itemize}

The serial ports are used to communicate with dedicated slow control instrumentation installed in the 
Detection Unit base and in the Calibration Unit.
The additional RS485 serial lines are required to trigger a laser emitter and an  acoustic emitter
used for the calibration of the \km3 detector; the acoustic interface is required
for a more sensitive device than the one inserted in each DOM, that will be used 
for scientific studies in different fields, including physics and marine biology.

\begin{figure}[tbp] 
\centering
\includegraphics[width=.6\textwidth]{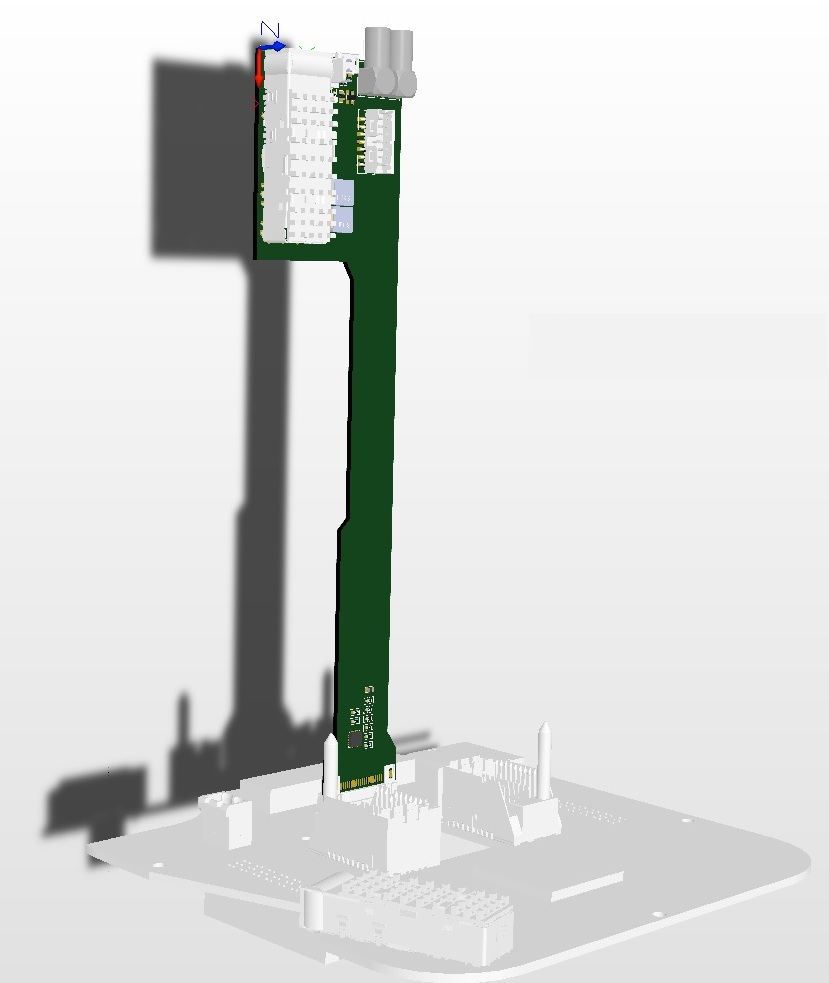}
\caption{The ``G'' Board plugged in the CLB; it  is used for testing and debugging 
during the assembly phase.}
\label{fig:tb}
\end{figure}

An additional auxiliary board (the so-called  ``G'' Board) have been designed   to allow   testing and
debugging of the CLB when it is assembled inside the full (or half) DOM. 
In this situation, in fact,
the space inside the DOM will be really crowded because of the presence of the 31~PMTs;
the CLB will not be accessible anymore, 
except using the optical line.
This could not be sufficient during the test phase, hence the ``G'' Board will be used, 
by inserting it with care in the DOM and plugging it to the CLB, as shown in figure~\ref{fig:tb}.

This board  allows the access to basic test functionalities  as:
\begin{itemize}
	\item JTAG chain
	\item USB ports
	\item LEMO connectors  for the timing synchronization  (Clock and Pulse Per Second signals)
	\item SFP connector to be used in case of problems with the one on the CLB
\end{itemize}

\section{Conclusions}

The Central Logic Board is responsible of the management of the Digital Optical
Module of the \km3 detector. It has to acquire data from 31~PMTs, from one acoustic device
and from other monitoring instrumentation, put them in UDP packets and send them
to the On-Shore station. A time sinchronization of the sub-nanosecond level
between all the \km3 DOMs is achieved by using White Rabbit.

The electronic group of the \km3 collaboration took care of the design, analysis
and test of the CLB and its auxiliary boards, needed to expand its
functionalities or for testing purposes.

Two series of prototypes were built, the first of 6~pieces and the second of 12~pieces.
Than a pre-serie production of 22~boards followed. Currently it is in production a 
new batch of 60~boards, and a mass production of 540~boards is expected by the end 
of the 2014.

\end{document}